\documentstyle[preprint,eqsecnum,pre,aps,oldlfont]{revtex}

\begin{document}
\draft
\title{
Critical Behavior of Period Doubling in Coupled Area-Preserving Maps
}
\author{Sang-Yoon Kim \cite{Kim}}
\address{
Department of Physics \\ Kangwon National University \\
Chunchon, Kangwon-Do 200-701, Korea
}
\maketitle
\begin{abstract}
We study the critical behavior of period doublings in $N$ symmetrically
coupled area-preserving maps for many-coupled cases with $N>3$.
It is found that the critical scaling behaviors depend on the
range of coupling interaction.
In the extreme long-range case of global coupling, in which each
area-preserving map is coupled to all the other area-preserving maps with
equal strength, there exist an infinite number of bifurcation routes in
the parameter plane, each of which ends at a critical point.
The critical behaviors, which vary depending on the type of bifurcation
routes, are the same as those for the previously-studied small $N$ cases
$(N=2,3)$, independently of $N$.
However, for any other non-global coupling cases of shorter range
couplings, there remains only one bifurcation route ending at the
zero-coupling critical point, at which the $N$
area-preserving maps become uncoupled,
The critical behavior at the zero-coupling point is also the same
as that for the small $N$ cases $(N=2,3)$, independently of the coupling
range.
\end{abstract}
%
%

\narrowtext

\section{Introduction}
\label{sec:INT}

Period doubling has been extensively studied in
area-preserving maps of McMillan form \cite{McMillan},
\begin{equation}
        \begin{array}{l}
         x(t+1) = - y(t) + f(x(t)), \\
         y(t+1) = x(t),
         \end{array}
\label{eq:2DM}
\end{equation}
where $(x(t),y(t))$ is a state vector at a discrete time t and $f$ is a
nonlinear function. A well-known example is the H${\acute {\rm e}}$non's
quadratic map
with $f(x) = 1 - ax^2$ \cite{Henon}. As the nonlinearity parameter $a$
increases, an initially stable orbit may lose its stability and give birth
to a stable period-doubled orbit. An infinite sequence of such
period-doubling bifurcations accumulates at a finite parameter value and
exhibits a universal asymptotic behavior.
However the asymptotic scaling behaviors for the area-preserving case
\cite{Benettin,Collet,Greene,Bountis,Helleman,Widom,MacKay}
are different from those for the one-dimensional dissipative case
\cite{Feigenbaum}.

An interesting question is whether the scaling results of
area-preserving maps extend to higher-dimensional volume-preserving maps.
Thus period doubling in four-dimensional volume-preserving
maps has attracted much interest in recent years
\cite{MacKay,Janssen,Mao1,Mao2,Mao3,Kim1}.
It has been found in Refs.~\cite{Mao2,Mao3,Kim1} that the critical
scaling
behaviors of period doublings for two symmetrically coupled area-preserving
maps are much richer than those for the uncoupled area-preserving case.
An infinite number of critical points form a critical Cantor set in the
space of the nonlinearity and coupling parameters. The critical behaviors
vary depending on the type of bifurcation routes to the critical points.
These scaling results hold also for the three-coupled case \cite{Mao4}.

It is also interesting to study whether or not the scaling behaviors for the
two- and three-coupled cases extend to arbitrary $N$-coupled $(N=2,3,\cdots)$
cases. Thus we study the critical behaviors of period doublings in
many coupled cases with $N>3$ and compare them with those
for the small $N$ cases $(N=2,3)$.
In Sec.~\ref{sec:CAPM} we introduce $N$ symmetrically
coupled area-preserving maps and discuss their general properties.
Stability of orbits in this $N$-coupled map is also discussed in
Sec.~\ref{sec:LS}. We study the critical scaling behaviors of period
doublings in Sec.~\ref{sec:CB}. It is found that the critical behaviors
for many-coupled cases with $N > 3$ depend on the coupling range.
In the extreme long-range case of global coupling,
the stable regions of in-phase orbits of period $2^n$
$(n=0,1,2,\cdots)$ form a ``bifurcation tree'' in the space of the
nonlinearity and coupling parameters, like the small $N$ cases $(N=2,3)$.
Consequently there exist an infinite number
of bifurcation routes in the parameter plane, each of which ends at a
critical point. The critical scaling behaviors, which vary depending on
the type of bifurcation routes, are also
the same as those for the small $N$ cases $(N=2,3)$, irrespectively of $N$.
However, for any other non-global coupling cases of
shorter range couplings only one bifurcation route ending at the
zero-coupling critical point is left in the parameter plane.
The critical scaling at the zero-coupling
critical point is also the same as that
for the small $N$ cases $(N=2,3)$, irrespectively of the coupling range.
Finally, a summary is given in Sec.~\ref{sec:SUM}.
%
%
\section{Coupled Area-Preserving Maps}
\label{sec:CAPM}

In this section we discuss general properties of $N$ coupled area-preserving
maps in which the coupling extends to the
$K$th [$1 \leq K \leq  {{\frac {N} {2}}}$  $({{\frac {N-1} {2}}})$ for even
(odd) $N$] neighbor(s) with equal strength.

Consider $N$ symmetrically coupled area-preserving maps with a periodic
boundary condition,
\begin{equation}
T:\left\{
        \begin{array}{l}
         {\vec{\bf x}}(t+1) = - {\vec{\bf y}}(t) + {\vec{\bf F}}({\vec{\bf x}}
         (t)), \\
         {\vec{\bf y}}(t+1) = {\vec{\bf x}}(t),
         \end{array}
   \right.
\label{eq:NCM1}
\end{equation}
where ${\vec{\bf x}} = (x_1,\dots,x_N)$, ${\vec{\bf y}} = (y_1,\dots,y_N)$,
${\vec{\bf F}}({\vec{\bf x}}) = (F_1({\vec{\bf x}}),\dots,F_N({\vec{\bf x}}))$,
and $N$ is a
positive integer larger than or equal to 2.
Here $z_m= (x_m,y_m)$ $(m=1,\dots,N)$ is the state vector of the $m$th
element of the $N$-coupled map $T$, and
the periodic boundary condition imposes
$z_m=z_{m+N}$ for all $m$.
The $m$th component
$F_m({\vec{\bf x}})$ of the vector-valued function ${\vec{\bf F}}
({\vec{\bf x}})$ is given by
\begin{eqnarray}
F_m({\vec{\bf x}}) &=& F(x_m,x_{m+1},\dots,x_{m-1}) \nonumber \\
        &=& f(x_m) + g(x_m,x_{m+1},\dots,x_{m-1}),
\end{eqnarray}
where $f$ is the nonlinear function of the uncoupled area-preserving
map (\ref{eq:2DM}), and $g$ is a coupling function obeying a condition
\begin{equation}
g(x,\dots,x) = 0\;\;{\rm for\;\;any}\;\;x.
\label{eq:CC}
\end{equation}
Thus the $N$-coupled map $T$ becomes:
\begin{equation}
T:\left\{
        \begin{array}{lll}
     x_m(t+1) = - y_m(t) &+& f(x_m(t)) \\
                         &+& g(x_m(t),\dots,x_{m-1}(t)), \\
     y_m(t+1) = x_m(t),&&m=1,\dots,N.
         \end{array}
   \right.
\label{eq:NCM2}
\end{equation}

This $N$-coupled map $T$ is called a symmetric map \cite{Mao2,Mao3,Kim1,Mao4}
since it has a cyclic
permutation symmetry such that
\begin{equation}
{\sigma}^{-1} T {\sigma} (\vec{\bf z}) = T (\vec{\bf z})\;\;\;{\rm for\;\;all}
\;\;\vec{\bf z},
\label{eq:CP}
\end{equation}
where $\vec{\bf z} = (z_1,\dots,z_N)$, $\sigma$ is a cyclic permutation
of $\vec{\bf z}$ such that
$\sigma {\vec{\bf z}} = (z_2,\dots,z_1)$, and $\sigma^{-1}$ is its
inverse.
The set of all fixed points of $\sigma$ is a two-dimensional (2D) subspace
of the $2N$
dimensional state space, on which
\begin{equation}
x_1=\cdots=x_N,\;\;y_1=\cdots=y_N.
\end{equation}
It follows from Eq.~(\ref{eq:CP}) that the cyclic permutation $\sigma$
commutes with the symmetric map $T$, i.e., $\sigma T = T \sigma$.
Hence
the 2D subspace becomes invariant under $T$, i.e., if a point $\vec {\bf z}$
lies on the 2D subspace, then its image $T {\vec {\bf z}}$ also lies on it.
An orbit is called an in-phase orbit if it lies on the 2D invariant subspace,
i.e., it satisfies
\begin{eqnarray}
x_1(t) &=& \cdots = x_N(t) \equiv x^{*}(t), \nonumber \\
y_1(t) &=& \cdots = y_N(t) \equiv y^{*}(t)\;\; {\rm for \;\;all}\;t.
\end{eqnarray}
Otherwise it is called an out-of-phase orbit.
Here we study only in-phase orbits. They can be easily found from the
uncoupled area-preserving map (\ref{eq:2DM}) since the coupling
function $g$ obeys the condition (\ref{eq:CC}).

The Jacobian matrix $DT$ of the $N$-coupled map $T$ is:
\begin{equation}
DT = \left( \begin{array}{cc}
              D{\vec{\bf F}} & \;\; -I_N \\
               I_N & \;\; 0
            \end{array}
     \right),
\end{equation}
where $D{\vec{\bf F}}$ is the Jacobian matrix of the function ${\vec{\bf F}}
({\vec{\bf x}})$,
$I_N$ is the $N \times N$ identity matrix, and 0 is the $N \times N$ null
matrix.
Since $Det(DT)=1$, the map $T$ is a $2N$ dimensional
volume-preserving map. Furthermore, if $D{\vec{\bf F}}$ is a symmetric matrix,
i.e.,
$D{\vec{\bf F}}^t=D{\vec{\bf F}}$ ($t$ denotes transpose), then the map $T$ is
a symplectic
map because its Jacobian matrix satisfies the relation
$DT^t\;J\;DT = J$ \cite{Howard}, where
\begin{equation}
J = \left( \begin{array}{cc}
              0 & \;\; I_N \\
               -I_N & \;\; 0
            \end{array}
     \right).
\end{equation}

The map $T$ is also reversible since it can be factored into the product
of two involutions \cite{Roberts}, $T = (TS)S$ [involutions satisfy
$(TS)^2=S^2=I$ (identity map)], where
\begin{eqnarray}
S &:& \left\{
        \begin{array}{l}
         {\vec{\bf x}}(t+1) = {\vec{\bf x}}(t), \\
         {\vec{\bf y}}(t+1) = - {\vec{\bf y}}(t) + {\vec{\bf F}}
         ({\vec{\bf x}}(t)),
         \end{array}
      \right. \\
TS &:& \left\{
        \begin{array}{l}
         {\vec{\bf x}}(t+1) = {\vec{\bf y}}(t), \\
         {\vec{\bf y}}(t+1) = {\vec{\bf x}}(t),
         \end{array}
   \right.
\end{eqnarray}
Such a decomposition represents a time-reversal symmetry because
the map $T$ is conjugate to its inverse $T^{-1}$ by the involution
$S$ such that $STS = T^{-1}$.
The sets of the fixed points of the involutions, called the symmetry
sets, are:
\begin{equation}
{\rm Fix}(S) =
\{ ({\vec{\bf x}},{\vec{\bf y}}) | {\vec{\bf y}}=
\frac {{\vec{\bf F}}({\vec{\bf x}})} {2} \},\;
{\rm Fix}(TS) =
 \{ ({\vec{\bf x}},{\vec{\bf y}}) | {\vec{\bf y}}={\vec{\bf x}} \}.
\end{equation}
An orbit is called a symmetric orbit if it is invariant under the involution
$S$; otherwise it is called an asymmetric orbit. It is easy to see that a
symmetric periodic orbit has two symmetric
points on the symmetry sets \cite{Roberts}: a symmetric orbit of even period
has two points on one symmetry set and none on the other, but a symmetric orbit
of odd period has one point on each symmetry set.
Here we study only symmetric periodic orbits.

Expressing the $N$-coupled map (\ref{eq:NCM2}) in the form of second-order
difference equations, we have:
\begin{eqnarray}
T: {x_m}(t+1) + {x_m}(t-1) &=& {F_m}({\vec{\bf x}}(t)) \nonumber \\
 &=& f(x_m(t)) \nonumber \\
        && + g(x_m(t),\dots,x_{m-1}(t)), \nonumber \\
&&m=1,\dots,N.
\label{eq:NCM3}
\end{eqnarray}
Consider an element, say the $m$th element, in this $N$-coupled map $T$.
Then the $(m \pm \delta)$th elements are called the $\delta$th neighbors
of the $m$th element, where $1 \leq \delta \leq {{\frac {N} {2}}}$
$({{\frac {N-1} {2}}})$ for even (odd) $N$.
If coupling extends to the $K$th neighbor(s), then the number $K$ is called
the range of the coupling interaction.

A general form of coupling for odd $N$ $(N \geq 3)$ is given by
\begin{eqnarray}
g(x_1,\dots,x_N) &=& {\frac {c} {2K+1}}
  {\sum_{l=-K}^{K}} [u(x_{1+l}) - u(x_1)], \nonumber \\
&=& c \left[ {\frac {1}{2K+1}} {\sum_{l=-K}^{K}} u(x_{1+l}) - u(x_1)
\right], \nonumber \\
&&K=1,\dots,{\frac {N-1} {2}},
\label{eq:GCF}
\end{eqnarray}
where $c$ is a coupling parameter and $u$ is a function of one
variable. Here the coupling extends to the $K$th neighbor(s) with
equal coupling strength, and the function $g$ satisfies the condition
(\ref{eq:CC}) because the value of $g$ for $x_1=\cdots=x_N$ becomes zero.
The extreme long-range interaction for
$K= {{\frac {N-1} {2}}}$ is
called a ``global'' coupling, for which the coupling function $g$ becomes
\begin{eqnarray}
g(x_1,\dots,x_N) &=& {\frac {c} {N}}{\sum_{m=1}^{N}} [u(x_{m}) - u(x_1)]
\nonumber \\
       &=& c \left[ {\frac {1}{N}} {\sum_{m=1}^{N}} u(x_m) - u(x_1) \right].
\label{eq:GC}
\end{eqnarray}
This is a kind of mean-field coupling, in which each element is coupled to
all the other elements with equal coupling strength. All the other couplings
with $K < {{\frac {N-1} {2}}}$ (e.g., nearest-neighbor
coupling with $K=1$) will be referred to as non-global couplings.
The $K=1$ case for $N=3$ corresponds to both the global coupling
and the nearest-neighbor coupling.

We next consider the case of even $N$ $(N \geq 2)$.
The form of coupling of Eq.~(\ref{eq:GCF}) holds for the cases of non-global
coupling with $K=1,\dots,{{\frac{N-2}{2}}}$ $(N \geq 4)$.
The global coupling for $K= {{\frac {N} {2}}}$ $(N \geq 2)$ also
has the form of Eq.~(\ref{eq:GC}),
but it cannot have the form of Eq.~(\ref{eq:GCF}) since there exists only one
farthest neighbor for $K= { {\frac{N}{2}} }$, unlike the odd
$N$ case.
The $K=1$ case for $N=2$ also corresponds to the nearest-neighbor coupling
as well as to the global coupling, like the $N=3$ case.
%
%
\section{Stability of in-phase orbits}
\label{sec:LS}

In this section we study stability of in-phase orbits in $N$ coupled
area-preserving maps.
In case of global coupling  the stability
region of an in-phase orbit in the parameter plane is the same
independently of $N$, whereas for the other non-global coupling cases of
shorter range couplings it depends on the coupling range $K$.

The stability analysis of an orbit in many coupled maps is
conveniently carried out by Fourier-transforming with respect to the
discrete space $\{m\}$ \cite{Kapral}. Consider an orbit
$\{ {\vec{\bf x}} (t) \} \equiv \{ {x_m}(t)\; ; \;m=1,\dots,N \}$ of the $N$
coupled map (\ref{eq:NCM3}).
The discrete spatial Fourier transform of the orbit is:
\begin{eqnarray}
{\cal F}[{x_m(t)}] &\equiv& {\frac{1}{N}} {\sum_{m=1}^{N}}
{e^{-2{\pi}imj/N}} {x_m}(t) = {\xi}_j(t), \nonumber \\
&&\;\;\;\;\;\;\;\;\;\; j=0,1,\dots,N-1.
\label{eq:FT}
\end{eqnarray}
The Fourier transform $\xi_j(t)$ satisfies $\xi_j^*(t) = \xi_{N-j}(t)$
($*$ denotes complex conjugate), and
the wavelength of a mode with index $j$ is ${\frac {N}{j}}$ for
$j \leq {{ {\frac {N} {2}}}}$ and
${\frac {N} {N-j}}$ for $j > {{\frac {N} {2}}}$.

To determine the stability of an in-phase orbit [$x_1(t) = \cdots =x_N(t)
\equiv x^{*}(t)$ for all $t$],
we consider an infinitesimal perturbation $\{ {\delta}x_m(t) \}$
to the in-phase orbit, i.e.,
$x_m(t)=x^{*}(t)+{\delta}x_m(t)$ for $m=1,\dots,N$.
Linearizing the $N$-coupled map
(\ref{eq:NCM3}) at the in-phase orbit, we obtain:
\begin{eqnarray}
{\delta}x_m(t+1) + {\delta}x_m(t-1) &=& f'(x^{*}(t))\;{\delta}x_m(t)
\nonumber \\
 && + {\sum_{l=1}^{N}} {G_l}(x^{*}(t))\; {\delta}x_{l+m-1}(t), \nonumber \\
 &&
\label{eq:LE}
\end{eqnarray}
where
\begin{eqnarray}
f'(x)={\frac{df}{dx}},\;\; \;{G_l}(x) & \equiv & \left. { \frac{\partial
g(\sigma^{(m-1)} {\vec{\bf x}})}{\partial x_{l+m-1}} }
\right |_{x_1=\cdots=x_N=x} \nonumber \\
&=& \left. { \frac{\partial g({\vec{\bf x}})}{\partial x_l} }
\right |_{x_1=\cdots=x_N=x}.
\label{eq:RCF}
\end{eqnarray}
Hereafter the functions $G_l$'s will be called ``reduced'' coupling functions
of $g({\vec{\bf x}})$.

Let ${\delta {\xi}_j}(t)$ be the Fourier transform of {$\delta x_m(t)$},
i.e.,
\begin{eqnarray}
\delta \xi_j (t)=
{\cal F}[{\delta x_m(t)}] &=& {\frac{1}{N}} {\sum_{m=1}^{N}}
{e^{-2{\pi}imj/N}} {\delta x_m}(t), \nonumber \\
&&\;\;\;\;\;\;\;\;j=0,1,\dots,N-1.
\end{eqnarray}
Then the Fourier transform of Eq.~(\ref{eq:LE}) becomes:
\begin{eqnarray}
\delta {\xi}_j(t+1) +\delta {\xi}_j(t-1) &=& [f'(x^{*}(t)) \nonumber \\
&&\; +{\sum_{l=1}^{N}} {G_l}(x^{*}(t))
{e^{2 \pi i(l-1)j/N}}] \nonumber \\
&& \times \, {\delta {\xi}_j}(t), \nonumber \\
&&\;j = 0,1,\dots,N-1.
\label{eq:LM1}
\end{eqnarray}
This equation can be also put into the following form:
\begin{eqnarray}
\left(
\begin{array}{l}
\delta{\xi}_j(t+1) \\
\delta{\xi}_j(t)
\end{array}
\right)
&=& L_j(t)
\left(
\begin{array}{l}
\delta{\xi}_j(t) \\
\delta{\xi}_j(t-1)
\end{array}
\right), \nonumber \\
&&\;\;\;\;\;\;\;\;j=0,1,\dots,N-1,
\label{eq:LM2}
\end{eqnarray}
where
\begin{equation}
L_j(t)=
\left( \begin{array}{cc}
f'(x^{*}(t))+{\displaystyle{\sum_{l=1}^{N}}} {G_l}(x^{*}(t))
{e^{2 \pi i(l-1)j/N}} & \;\; -1 \\
1 &\;\; 0
        \end{array}
\right).
\label{eq:JML}
\end{equation}
Note that the determinant of $L_j$ is one, i.e., $Det(L_j)=1$.

Stability of an in-phase orbit of period $q$ is
determined by iterating Eq.~(\ref{eq:LM2}) $q$ times:
\begin{eqnarray}
\left(
\begin{array}{l}
\delta{\xi}_j(t+q) \\
\delta{\xi}_j(t+q-1)
\end{array}
\right)
&=& M_j
\left(
\begin{array}{l}
\delta{\xi}_j(t) \\
\delta{\xi}_j(t-1)
\end{array}
\right), \nonumber \\
&&\;\;\;\;\;\;\;\;j=0,1,\dots,N-1,
\end{eqnarray}
where
\begin{equation}
M_j= {\prod_{k=t}^{t+q-1}} L_j(k).
\end{equation}
That is, the stability of each mode with index $j$ is
determined by the $2 \times 2$ matrix $M_j$.
Since $Det(M_j)=1$, each matrix $M_j$ has a reciprocal
pair of eigenvalues, $\lambda_j$ and $\lambda_j^{-1}$.
These eigenvalues are called the stability multipliers of the mode with
index $j$. We also
associate with a pair of multipliers $(\lambda_j,\lambda_j^{-1})$
a stability index,
\begin{equation}
\rho_j = \lambda_j + \lambda_j^{-1},\;\;j=0,1,\dots,N-1,
\end{equation}
which is just the trace of $M_j$, i.e., $\rho_j=Tr(M_j)$.

It follows from the condition (\ref{eq:CC}) that the
reduced coupling functions satisfy
\begin{equation}
{\sum_{l=1}^{N}} G_l(x) = 0.
\end{equation}
Hence the matrix of Eq.~(\ref{eq:JML}) for $j=0$ becomes:
\begin{equation}
L_0(t)=
\left( \begin{array}{cc}
f'(x^{*}(t)) & \;\; -1 \\
1 &\;\; 0
        \end{array}
\right).
\end{equation}
This is just the Jacobian matrix of the uncoupled area-preserving map.
Therefore the stability index $\rho_0$ of the $j=0$ mode is just
that for the case of the area-preserving map, i.e., $\rho_0$ depends
only on the nonlinearity parameter $a$.
While there is no coupling
effect on $\rho_0$, coupling generally affects the other stability
indices $\rho_j$'s ($j \neq 0$).

In case of the global coupling of Eq.~(\ref{eq:GC}),
the reduced coupling functions become:
\begin{equation}
{G_l}(x) = \left \{
 \begin{array}{l}
  (1-N) G(x)\;\;\;\; {\rm for}\;l=1, \\
  \;\;\;\;\;\;G(x)\;\;\;\;\;\;\;\;\;\;{\rm for}\;l \neq 1,
  \end{array}
  \right.
\end{equation}
where $G(x)= {{\frac{c}{N}}} u'(x)$.
Substituting $G_l$'s into the $(1,1)$ entry of the matrix $L_j(t)$, we
have:
\begin{equation}
{\sum_{l=1}^{N}} G_l(x) e^{2 \pi i(l-1)j/N} =
\left \{ \begin{array}{l}
          \;\;\;\;\;0\;\;\;\;\;\;\;{\rm for}\;\; j=0, \\
          -c\, u'(x)\;\;{\rm for}\;\; j \neq 0.
         \end{array}
\right.
\label{eq:SE2}
\end{equation}
Hence all stability indices $\rho_j$'s of modes with non-zero index $j$
$(j \neq 0)$ become real and the same, i.e., $\rho_1 = \cdots = \rho_{N-1}$.
Thus there exist only two independent real stability indices $\rho_0$ and
$\rho_1$, the values of which are also independent of $N$.

We next consider the non-global coupling of the form (\ref{eq:GCF}) and
define
\begin{equation}
G(x) \equiv {\frac {c} {2K+1}} u'(x),
\end{equation}
where $1 \leq K \leq {{{\frac {N-2} {2}}}}\;
({{\frac {N-3} {2}}})$ for even (odd) $N$ larger than 3.
Then we have
\begin{equation}
{G_l}(x) = \left \{
 \begin{array}{l}
  -2 K G(x)\;\;\; {\rm for}\;l=1, \\
  \;\;\;\;\;G(x)\;\;\;\;\;\; {\rm for}\; 2 \leq l \leq 1+K \;\; {\rm or} \\
  \;\;\;\;\;\;\;\;\;\;\;\;\;\;\;\;\;\;\;{\rm for}\;N+1-K \leq l \leq N, \\
  \;\;\;\;\;\;\;0\;\;\;\;\;\;\;\;\;\;{\rm otherwise.}
  \end{array}
  \right.
\end{equation}
Substituting the reduced coupling functions into
the matrix $L_j(t)$ of Eq.~(\ref{eq:JML}), the second term of the $(1,1)$
entry of $L_j(t)$ becomes:
\begin{equation}
{\sum_{l=1}^{N}} G_l(x) e^{2 \pi i(l-1)j/N}
= - {S_N}(K,j) c\, u'(x),
\label{eq:SE1}
\end{equation}
where
\begin{equation}
{S_N}(K,j) \equiv {4 \over {2K+1}} {\sum_{k=1}^{K}}
sin^2 {{\pi jk} \over {N}}
= 1- {\frac {sin(2K+1) {{\frac{\pi j}{N}}}}
{(2K+1) sin{{\frac{\pi j}{N}}}}}.
\label{eq:SF}
\end{equation}
Hence all stability indices $\rho_j$'s with non-zero index $(j \neq 0)$
become real, but they vary depending on the coupling range $K$ as well as
on the mode number $j$. Since $S_N(K,j) = S_N(K,N-j)$, the real stability
indices satisfy
\begin{equation}
\rho_j = \rho_{N-j},\;\;j=0,1,\dots,N-1.
\end{equation}
Thus it is sufficient to consider only the case of
$0 \leq j \leq {N \over 2}$ $({{N-1} \over 2})$ for even (odd) $N$.
Comparing the expression in Eq.~(\ref{eq:SE1}) with that in
Eq.~(\ref{eq:SE2}) for $j \neq 0$, one can easily see that
they are the same except for the factor $S_N (K,j)$. Consequently,
making a change of the coupling parameter
${c \rightarrow {c \over {S_N (K,j)}}}$, the stability index
$\rho_j$ for the non-global coupling case of range $K$ becomes the same as
that for the global-coupling case.

It follows from the reality of $\rho_j$ that the reciprocal pair of
eigenvalues of $M_j$ lies either on the
unit circle, or on the real line in the complex plane, i.e., they are
a complex conjugate pair on the unit circle, or a reciprocal pair of reals.
Each mode with index $j$ is stable if and only if the magnitude of its
stability index $\rho_j$ is less than or equal to two ($|\rho_j| \leq 2$),
i.e., its stability multipliers are a pair of complex conjugate numbers
of modulus unity.
A period-doubling (tangent) bifurcation occurs when
the stability index $\rho_j$ decreases (increases) through $-2$ (2), i.e.,
two eigenvalues coalesce
at $\lambda_j = -1$ (1) and split along the negative (positive) real axis.

When the stability index $\rho_0$ for an in-phase orbit in an $N$-coupled map
decreases
through $-2$, the in-phase orbit loses its stability via in-phase
period-doubling bifurcation, giving rise to the birth of the period-doubled
in-phase orbit. Here we are interested in such in-phase period-doubling
bifurcations.
Thus, for each mode with non-zero index $j$ $(j \neq 0)$ we consider a region
in the space of the nonlinearity and coupling parameters, in whcih both
modes with indices $0$ and $j$ are stable. This stable region is bounded by
four bifurcation lines of both modes (i.e., those curves determined by the
equations $\rho_0 = \pm 2$ and $\rho_j = \pm 2$), and it will be denoted by
$U_N$.

In case of global coupling, those stable regions coincide, independently
of $N$ and $j$, because all stability indices $\rho_j$'s of modes with
non-zero $j$ are the same, irrespectively of $N$; the stable region for this
case will be denoted by $U_G$.
An in-phase orbit is stable only when all its modes are stable. For the
global-coupling case, $U_G$ itself is the stability region of the in-phase
orbit irrespectively of $N$, because all modes are stable in the region
$U_G$.

However the stable regions $U_N$'s vary depending
on the coupling range $K$ and the mode number $j$ for the non-global coupling
cases, i.e., $U_N = U_N (K,j)$. To find the stability region of an in-phase
orbit in an $N$-coupled map with a given $K$, one may start with the
stability region $U_G$ for the global-coupling case.
Rescaling the coupling parameter $c$ by a scaling factor
$ {1 \over {S_N (K,j)}}$ for each
non-zero $j$ $(j \neq 0)$, the stable region $U_G$ is transformed into a
stable region $U_N (K,j)$. Then the stability region of the in-phase orbit
is given by the intersection of all such stable regions $U_N$'s.
%
%
\section{CRITICAL BEHAVIOR OF PERIOD DOUBLINGS}
\label{sec:CB}

We are concerned about the critical scaling behavior of period doublings
of in-phase orbits in $N$ symmetrically coupled area-preserving maps
(\ref{eq:NCM3}).
Small $N$ cases $(N=2,3)$ have been studied in
Refs.~\cite{Mao1,Mao2,Mao3,Kim1,Mao4}.
It is interesting to study whether or not the critical scalings for
the small $N$ cases extend to the large $N$ cases.
Thus we study the critical behaviors in the large $N$ cases
and find that those for $N >3$ depend on the coupling range $K$.
For the $N$-coupled map (\ref{eq:NCM3}),
we choose $f(x) = 1 - a x^2$ as the nonlinear
function of the uncoupled area-preserving map and consider separately
two kinds of couplings, the global and non-global coupling cases.

\subsection{Global Coupling}
\label{subsec:GC}
We first study an $N$-coupled map with global coupling, in which each
element is coupled to all the other elements with equal strength.
As shown in Sec.~\ref{sec:LS}, all stability indices $\rho_j$'s of modes
with non-zero $j$ are not only the same (i.e., $\rho_1 = \cdots
=\rho_{N-1}$), but also independent of $N$ for the global-coupling case.
Thus the stability diagram of in-phase orbits in the space
of the nonlinearity and coupling parameters becomes the same as that
for the two-coupled case, independently of $N$. That is, the stable regions
of in-phase orbits of period $2^n$ $(n=0,1,2,\cdots)$ form a ``bifurcation
tree'' in the parameter plane. Consequently there exist an infinite number of
bifurcation routes in the parameter plane, each of which ends at a critical
point.
The critical behaviors, which vary depending on the type of the bifurcation
routes, are also the same as those for the two-coupled case, irrespectively
of $N$ (for details of the $N=2$ case, refer to Refs.~\cite{Mao2,Mao3,Kim1}).

As an example, we consider a linearly-coupled case in which the coupling
function (\ref{eq:GC}) is
\begin{equation}
g(x_1,\dots,x_N)
       = c \left[ {\frac {1}{N}} {\sum_{m=1}^{N}} x_m - x_1 \right].
\label{eq:LC}
\end{equation}
Figure \ref{figure1} shows the stability diagram of in-phase orbits
with period
$q=1,\,2,\,4$. These low-period orbits can be analytically obtained from the
uncoupled second-order difference equation, $x(t+1) + x(t-1) = f(x(t))$
\cite{Bountis}.

The period-1 fixed point is
\begin{equation}
x^* (0) = {1 \over a} (-1 + \sqrt{1+a} ),
\label{eq:FP}
\end{equation}
and the independent stability indices are
\begin{eqnarray}
&\rho_0& (a)= 2 (1- \sqrt{1+a}),\\ &\rho_1& (a,c)= \rho_0 (a) - c.
\end{eqnarray}
Its stable region in the parameter plane
is limited by four bifurcation lines associated with tangent and
period-doubling bifurcations of both modes (i.e., those curves determined
by the equations $\rho_j = \pm 2$ for $j =0,1$).

When the stability index $\rho_0$ of the fixed point decreases through
$-2$, it loses its stability via in-phase period-doubling bifurcations,
and gives rise to the birth of an in-phase orbit of period 2,
\begin{equation}
x^* (0) = {1 \over a} (1- \sqrt{a-3}),\;\;\;x^* (1) = {1 \over a}
(1+ \sqrt{a-3}).
\end{equation}
Two independent stability indices of the period-2 orbit are
\begin{eqnarray}
&\rho_0& (a) = -4a + 14,\\ &\rho_1& (a,c) = \rho_0 (a) +c (c+4).
\end{eqnarray}
Note that two period-2 stable branches bifurcate out of the
period-1 stable region in the parameter plane (see Fig.~\ref{figure1}).

An in-phase orbit of period 4,
\begin{eqnarray}
x^* (0) &=& \left( { {\sqrt{a} - 2} \over {a \sqrt{a}} } \right)^{1 \over 2},
\;\;x^* (1) = {1 \over \sqrt{a}}, \nonumber \\
x^* (2) &=& - x^* (0),\;\;x^* (3) =
x^* (1),
\end{eqnarray}
bifurcates from the in-phase period-2 orbit at $a=4$. Its two independent
stability indices are
\begin{eqnarray}
\rho_0 (a) &=& 16 a (2 \sqrt{a} - a) +2,  \\
\rho_1(a,c) &=& \rho_0 (a) +c^4 +4 \sqrt{a} c^3
         + 4 (2 \sqrt{a} -1 ) c^2  \nonumber \\
       && - 8 \sqrt{a} (2a - 4 \sqrt{a}+1) c.
\end{eqnarray}
As shown in Fig.~\ref{figure1}, each of the two period-2 stable branches
also bifurcates
into two period-4 branches. Thus the stability region for this case consists
of 4 stable branches.

However it is difficult to obtain analytically  higher periodic orbits with
period larger than four. Thus one usually resorts to numerical computation.
Successive bifurcations of stability regions have been also observed
for the case of higher periods \cite{Mao2,Kim1}. That is, each
``mother'' stable branch bifurcates into two ``daughter'' stable branches
successively in the parameter plane; hereafter we call the direction of the
left (right) branch of the two daughter branches $L$ $(R)$ direction.
Consequently the stability region of the in-phase orbit of period $2^n$
$(n=0,1,2,\dots)$ consists of $2^n$ branches. Each branch can be represented
by its address $[a_0 , \dots , a_n ]$, which is a sequence of symbols
$L$ and $R$ such that $a_0 = L$ and $a_i = L$ or $R$ for $i \geq 1$.

An infinite sequence of connected stable branches (with increasing period)
is called a bifurcation ``route'' \cite{Mao2,Kim1}. Each bifurcation route
is also represented by an infinite sequence of symbols $L$ and $R$. Hence
there exist an infinite number of bifurcation routes in the parameter plane.
A bifurcation ``path'' in a bifurcation route is formed
by following a sequence of parameters $(a_n , c_n )$, at which the in-phase
orbit of level $n$ (period $2^n$) has some given stability indices
$(\rho_0 , \rho_1 )$ \cite{Mao2,Kim1}. All bifurcation paths within a
bifurcation route converge to an accumulation point $(a^* , c^*)$. Here the
value of $a^*$ is always the same as that of the accumulation point for
the area-preserving case (i.e., $a^*=4.136\,166\,803\,904\dots$), whereas
the value of $c^*$ varies depending on the bifurcation routes. Thus each
bifurcation route ends at a critical point $(a^* , c^*)$ in the parameter
plane.

The nonlinearity-parameter values $a_n$ geometrically converge to
a limit value $a^*$ in the limit of large $n$, i.e.,
\begin{equation}
a_n - a^* \sim \delta_1^{-n} \;\;\;{\rm for\;large\;} n,
\end{equation}
where the value of $\delta_1$ is always the same as that of the scaling
factor
$\delta$ $(=8.721\dots)$ for the uncoupled area-preserving case,
independently of bifurcation routes.
However the scaling behaviors associated with the coupling parameter
depend on the type of bifurcation routes.

Consider $p$ $(p=1,2,\dots)$ subsequences $\{ c_{pm+j};\;m=0,1,2,\dots \}$
$(j=0,\dots,p-1)$ of the coupling-parameter sequence
$\{ c_n \}$.
Then a bifurcation route is called a ``period-$p$'' route \cite{Kim1}
if the $p$ subsequences geometrically converge to a limit value $c^*$
in the limit of large $m$, i.e.,
\begin{equation}
c_{pm+j} - c^* \sim \delta_2^{-m} \;\;\;{\rm for\;large\;} m,
\end{equation}
where the value of $\delta_2$ is independent of $j$, but it depends
on the type of bifurcation routes.

As an example, consider the case of the lowest period-1 bifurcation routes
\cite{Mao2,Mao3,Kim1}.
There exist three kinds of period-1 routes, called the $S$, $A$, and $E$
routes. An $S$ route is followed if one goes asymptotically in the same
direction ($L$ or $R$ direction). Hence its address has the form,
$[u,(L,)^\infty ]$ $(\equiv [u,L,L,\dots])$ or $[v,(R,)^\infty ]$,
where $u$ and $v$ are arbitrary finite sequences of $L$ and $R$.
The coupling-parameter scaling factor for this case is
$\delta_2=4.000$. Unlike the case of $S$ routes, the direction
of an $A$ route asymptotically alternates between the $L$ direction
and the $R$ direction. Hence its address has the form,
$[w,(L,R,)^\infty ]$, where $w$ is an arbitrary finite sequence of $L$ and
$R$. The value of $\delta_2$ is $-2.000$ for all $A$ routes except for
the case of one special route, called the $E$ route. The address of the
$E$ route is $[(L,R,)^\infty ]$ (i.e., the sequence $w$ for this case is
empty). (Hereafter only the routes with non-empty $w$ sequences will be
called
$A$ routes.) The limit value of the coupling parameter for the $E$ route
is $c^* =0$, and the scaling factor is $\delta_2 = -4.404$.

In order to see the periodicity of a bifurcation route, a
``route sequence'' has been introduced in Ref.~\cite{Kim1}. It is uniquely
determined by the address of the route as follows.
If the $n$th element of the address is the same as the next $(n+1)$th
element, we assign a $0$ to the $n$th element of the route sequence;
otherwise we assign a $1$. Thus the route sequence becomes an infinite
sequence of two symbols $0$ and $1$.

As an example we examine the asymptotic patterns of route sequences for
the case of the period-1 routes.
The route sequence of an $S$ route is $[u',(0,)^\infty ]$ because one goes
asymptotically in the same direction in an $S$ route; here $u'$ is an
arbitrary finite sequence of $0$ and $1$. On the other hand, the route
sequence of an $A$ or the $E$ route is $[w',(1,)^\infty ]$ because
the direction of an $A$ or the $E$ route asymptotically alternates between
the $L$ and $R$ directions. Here $w'$ is an arbitrary non-empty sequence of
$0$ and $1$ for the case of $A$ routes, whereas it is empty for the $E$
route. Note that the
route sequence of any period-1 route exhibits eventually a period-1 pattern.

{}From the asymptotic period-$p$ patterns of the route sequences for the
low-period $(p=1,2)$ routes, it has been conjectured in \cite{Kim1} that
if the route sequence of a bifurcation route exhibits an asymptotic
period-$p$ ($p=1,2,\dots$) pattern, the route becomes a
period-$p$ one; otherwise it becomes a random one in which the
coupling-parameter values $c_n$ randomly converge to a limit value
$c^*$ without any periodicity.

\subsection{Non-global Coupling}
\label{subsec:NGC}

Here we study the non-global coupling cases with
$K < {\frac {N} {2}}$ $({\frac {N-1} {2}})$ for even (odd) $N$ larger than
three.
Of the infinite kinds of bifurcation routes for the global coupling case,
only the $E$ route ending at the zero-coupling critical point $(a^*,0)$ is
left as a bifurcation route in the parameter plane for all the non-global
coupling cases. The critcal scaling behaviors near the zero-coupling
critical point are also the same as those for the $N=2$ and $3$ cases of
global coupling.

As an example we consider a linearly-coupled, nearest-neighbor coupling
case with $K=1$, in which  the coupling function is
\begin{equation}
g(x_1,\dots,x_N) = \left \{ \begin{array}{c}
        {{\frac {c} {2}}}\; (x_2 - x_1)\;\; {\rm for}\;\;N=2, \\
    {{\frac {c} {3}}}\; (x_2 + x_N-2 x_1)\;\;{\rm for}\;\;
     N \geq 3.
                             \end{array}
                   \right.
\end{equation}
Note that the two- and three-coupled maps for $N=2,\,3$ correspond to the
global coupling case.
The reduced coupling functions $G_l$'s $(l=1,\dots,N)$ of $g$ defined in
Eq.~(\ref{eq:RCF})
are given by
\begin{eqnarray}
G_2(x) &=& {\frac {c}{2}} \equiv G(x),\;\;G_1(x) = -G(x)\;\;
{\rm for}\;\;N=2, \nonumber \\
&& \\
G_2(x) &=& G_N(x) = {\frac {c}{3}} \equiv G(x), \nonumber \\
G_1(x) &=& -2 G(x),\;\;
G_l(x)=0 \;\;(l \neq 1,2,N)\;\;{\rm for}\;\;N \geq 3. \nonumber \\
&&
\end{eqnarray}
Substituting $G_l$'s into the (1,1) entry of the matrix $L_j(t)$
of Eq.~(\ref{eq:JML}), we obtain:
\begin{equation}
{\sum_{l=1}^{N}} G_l(x^{*}(t)) e^{2 \pi i(l-1)j/N} =
\left \{ \begin{array}{c}
          -c\, \delta_{j1}\;\;{\rm for}\;\; N=2,\,3, \\
          - S_N (1,j) c \;\;   {\rm for}\;\; N > 3,
         \end{array}
\right.
\end{equation}
where $\delta_{j1}$ is the Kronecker delta (i.e., $\delta_{j1}$ is
0 for $j=0$ and 1 for $j=1$), and $S_N (1,j) = { {4 \over 3}
sin^2 {{\pi j} \over N} }$.
Making a change of the coupling parameter
${c \rightarrow {c \over {S_N (1,j)}}}$ for each non-zero $j$,
the stability index
$\rho_j$ for $N>3$ becomes the same as that for the $N=2$ and $3$ cases
of global coupling.

For each mode with non-zero index $j$, consider the stability region
$U_N (1,j)$, in which both modes with indices $0$ and $j$ are stable.
The stability region $U_G$ for the global-coupling case is independent of
both $N$ and $j$ because all stability indices $\rho_j$'s of modes with
non-zero $j$ are the same, independently of $N$.
Rescaling the coupling parameter $c$ with the scaling factor
$ {1 \over S_N (1,j)}$,
the stability region $U_G$ for the $N=2$ and $3$ cases of global
coupling is transformed into the stablity region $U_N (1,j)$ for $N>3$.
Since the factor $S_N$ is dependent on the index $j$, the stability region
$U_N (1,j)$ varies depending on the index $j$.
Then the stability region of an in-phase orbit, in which all its modes
are stable, is given by the intersection of all such
stable regions $U_N$'s.

As an example we consider the case $N=4$.
Figure \ref{figure2} shows the stability regions of the $2^n$-periodic
$(n=1,2,3,4)$ orbits for this case.
We first note that the scaling factor ${1 \over S_4 (1,j)}$
has its minimum value ${\frac {3} {4}}$
at $j=2$. However $U_4 (1,2)$ itself cannot be the stability region
of the in-phase orbit of level 1 (i.e., $n=1$), because bifurcation curves
of different modes with non-zero indices intersect one another.
As shown in Fig.~\ref{figure2}(a),
the branch including a $c=0$ line segment remains unchanged,
whereas the other branch becomes flattened by the bifurcation curve of the
mode with $j=1$.
Due to the successive flattening with increasing level $n$
[see Figs.~\ref{figure2}(a) and (b)],
of the infinite kinds of bifurcation routes for the
global-coupling case remains only the $E$ route ending at the zero-coupling
point $(a^* , 0)$. Thus only the zero-coupling point is left as a
critical point in the parameter plane.
Note also that Figs.~\ref{figure2}(a) and (b) nearly coincide near the
zero-coupling point except for small numerical differences.

We now examine the scaling behaviors near the zero-coupling critical point
for the $N=4$ case of nearest-neighbor coupling.
Consider a self-similar sequence of parameters $(a_n , c_n )$, at which
the in-phase orbit of level $n$ has some given stability indices, in the
$E$ route for the global-coupling case. Rescaling the coupling parameter
with the factor $ {3 \over 4}$,
this sequence is transformed into a self-similar one for the $N=4$ case
of nearest-neighbor coupling.
Thus the ``width'' of each stability region in the $E$ route for the case
of the global coupling is reduced to that for the $N=4$ case of
nearest-neighbor coupling by the scaling factor $ {3 \over 4}$, whereas
the ``heights'' of all stable regions in the $E$ route remain unchanged
\cite{Kim2}. It is therefore obvious that the scaling
behaviors near the zero-coupling critical point for the nearest-neighbor
coupling case are the same as those for the global coupling case. That is,
the height and width $h_n$ and $w_n$ of the stability region of level $n$
geometrically contract in the limit of large $n$,
\begin{equation}
h_n \sim \delta_1^{-n},\;\;w_n \sim \delta_2^{-n}\;\;{\rm for\;large\;}
n,
\end{equation}
where $\delta_1 = 8.721\dots$ and $\delta_2 = -4.404$.

The scaling results for the nearest-neighbor coupling case
with $K=1$ extend to all the other non-global coupling cases with
$1 < K < {{N \over 2}}\; ({{N-1 \over 2}})$ for even (odd) $N$.
For each non-global coupling case with $K>1$, consider a mode with
index $j_{\rm max}$ for which the
factor $S_N (K,j)$ of Eq.~(\ref{eq:SF}) becomes the largest one and
its stability region $U_N (K,j_{\rm max})$ including a $c=0$ line segment.
Here the value of $j_{\rm max}$ varies depending on the range $K$.
Like the nearest-neighbor coupling case with $K=1$, of the two stable
branches the one including the $c=0$ line segment remains unchanged,
whereas the other one becomes flattened by the bifurcation curves of the
other modes with non-zero indices. Thus the overall shape of the stability
diagram in the parameter plane becomes essentially the same as that for the
case with $K=1$. Consequently  only the $E$-route ending at the zero-coupling
point is left as a bifurcation route, and the scaling behaviors near
the zero-coupling critical point are also the same as those for the
global-coupling case.
%
%

\section{Summary}
\label{sec:SUM}

The critical behaviors of period doublings in $N$ symmetrically coupled
area-preserving maps are studied for many-coupled cases with
$N>3$. It is found that the critical scaling behaviors depend
on the coupling range $K$. For the global-coupling case
[$K={N \over 2}\;({N-1 \over 2})$ for even (odd) $N$], the stable regions
of in-phase orbits with period $q=2^n$
$(n=0,1,2,\dots)$ form a ``bifurcation tree'' in the space of the
nonlinearity and coupling parameters, like the small $N$ cases $(N=2,3)$.
Hence there exist an infinite number of bifurcation routes in the parameter
plane, each of which ends at a critical point. The
critical behaviors, which vary depending on the type of the bifurcation
routes, are also the same as those
for the small $N$ cases $(N=2,\;3)$, independently of $N$.
However, for any other non-global coupling cases
[$1 \leq K < {N \over 2}\;({N-1 \over 2})$ for even (odd) $N$],
only the $E$-route ending at the zero-coupling point is
left as a bifurcation route in the parameter plane. The critical
behaviors at the zero-coupling critical point are also the same as those
for the small $N$ cases $(N=2,3)$, independently of $K$.

%
%

\acknowledgments
This work was supported by the Basic Science Research Institute
Program, Ministry of Education, Korea, 1994, Project No. BSRI-94-2401.

%
%

\begin{figure}
\caption{
Stability  diagram of in-phase orbits in $N$ linearly coupled
maps with the global coupling. The horizontal (non-horizontal) solid and
dashed lines denote the period-doubling and tangent bifurcation lines
of the $j=0$ $(1)$ mode, respectively. Note that all stability indices
$\rho_j$'s of modes with non-zero $j$ are the same (i.e., $\rho_1 =
\cdots = \rho_{N-1}$) for the global-coupling case.
For other details see the text.
   }
\label{figure1}
\end{figure}

\begin{figure}
\caption{
Stable regions of the in-phase period-$2^n$ orbits of level $n=1,\,2,\,3,\,4$
in four linearly coupled maps with the nearest neighbor coupling. The case
of $n=1$ and $2$ $(3$ and $4$) is shown in (a) [(b)].
Each stable region is bounded by its solid boundary curves.
The period-doubling (tangent) bifurcation curve of the $j$
mode of an in-phase orbit with period q is denoted by a symbol $q_j^{p(t)}$.
The values of the scaling factors used in (b) are $\delta_1=8.721$ and
$\delta_2=-4.404$.
}
\label{figure2}
\end{figure}

%
%


\begin{references}
\bibitem{Kim2} We define the ``width'' $w_n$ of the stable region of level
$n$ in the $E$ route as the length of
the boundary line segment associated with the period-doubling
bifurcation of the $j=0$ mode. The ``height'' $h_n$ is also defined as the
length of the $c=0$ line segment inside the stable region.
\end{references}
\end{document}